\documentclass[preprint,aps,prl,showpacs]{revtex4}

\begin{document}
\title{Experimental Study of Positionally Disordered Josephson Junctions Arrays}
\author{Young-Je Yun}
\author{In-Cheol Baek}
\author{Mu-Yong Choi}
\affiliation{BK21 Physics Division and Institute of Basic Science,
Sungkyunkwan University, Suwon 440-746, Korea}

\begin{abstract}
We experimentally studied the effect of positional disorder on a
Josephson junction array with $f = n$, $\frac{1}{2} + n$, or
$\frac{2}{5} + n$ flux quanta per unit cell for integral $n$.
This system provides an experimental realization of a
two-dimensional $XY$ model with random phase shifts. Contrary to
many earlier numerical and analytical investigations, our results
suggest that low-temperature superconductivity is never destroyed
by positional disorder. As the disorder strength increased, the
Kosterltz-Thouless (KT) type order in the $f = 0$ and
$\frac{1}{2}$ systems changed to a non-KT type order with a
long-range phase coherence, which persisted even in the maximal
disorder limit. A possible finite-temperature glass transition is
discussed.
\end{abstract}


\maketitle

The $XY$ model with quenched random phase shifts, also known as
the random-gauge $XY$ model, has been intensively studied with a
focus on the maximal disorder limit as a model for a vortex glass
phase \cite{R1} of high-temperature superconductors in a magnetic
field. Despite a great deal of effort for the last two decades,
the model in two dimensions (2D) remains not well understood.
Conflicting evidences, mostly numerical or analytical ones,
forbid a consensus about the possibility of a finite-temperature
glass transition in 2D. The purpose of this work is to see
experimentally how the low-temperature ordered phase of the 2D
random-gauge $XY$ system evolves with the disorder strength
increased.

The random-gauge $XY$ model is described by the Hamiltonian
\begin{equation} H = - J\sum\limits_{ij} {\cos (\theta _i  - \theta _j  -
A_{ij} )}, \end{equation} where $\theta _i $ is the phase of the
order parameter on site $i$ and the quenched random phase shifts
$A_{ij} $'s are uncorrelated and distributed uniformly in
$[A_{ij}^o  - r\pi ,A_{ij}^o  + r\pi ]$ with $0 \le r \le 1$. The
case with maximal disorder corresponds to the disorder strength
$r = 1$. The effect of random phase shifts in 2D was investigated
primarily with respect to the Kosterlitz-Thouless (KT) transition
for $f = 0$, where frustration $f = (1/2\pi )\sum {A_{ij}^0 } $.
The directed sum of the $A_{ij}^0 $ is around a unit cell. A
prevailing conclusion in many numerical and analytical studies
\cite{R2} for $f = 0$ is that there exists a KT transition at
finite temperatures when $r < r_c  \approx 0.37$ and the case
with maximal disorder, termed the gauge glass, has no ordered
phase at any finite temperature. The lack of ordering at finite
temperatures has been proven \cite{R3} rigorously in the limit of
maximum disorder. Nevertheless, numerical evidences against this
conclusion have been also provided by several different research
groups \cite{R4,R5}. The counter evidences point to a
finite-temperature superconducting glass order in the gauge glass
limit. It has been shown \cite{R6} numerically that a breaking of
ergodicity due to the large energy barrier against vortex motion
may open up the possibility of a finite-temperature glass
transition which was denied by the analytical  equilibrium
studies \cite{R3}. The confused situation is much the same for $f
= \frac{1}{2}$, where the $XY$ system without disorder has the
Ising-like structural order as well as the KT-like phase
coherence at low temperatures \cite{R7}. Along with arguments
\cite{R8} for the persistence of both the Ising-like order and
the KT-like order for at least weak disorder strength, it has
been also suggested \cite{R9} that the critical disorder for $f =
\frac{1}{2}$ is zero, or that any amount of disorder, no matter
how weak it is, destroys the Ising-like order and the phase
coherence even at zero temperature.

The disagreement among the analytical and numerical results may
be settled by experimental evidences for the random-gauge effect.
The random-gauge $XY$ model in 2D can be most closely realized by
a Josephson junction array (JJA) with positional disorder in a
uniform magnetic field \cite{R10}. For a JJA, $J$ is the Josephson
coupling energy per junction and given by $J = \hbar i_c /2e$
with $i_c $ the single-junction critical current. $\theta _i $ is
the phase of the superconducting island at site $i$ and $A_{ij} =
(2\pi /\phi _0 )\int_i^j {\mathbf{A}  \cdot d\mathbf{l} } $ is
the integral of the vector potential across a junction with $\phi
_0 $ being the flux quantum. The frustration $f$ is equal to the
mean number of flux quanta penetrating each cell. Random phase
shifts can be introduced by randomly displacing the centers of
superconducting islands while keeping the junctions on their
regular positions in a periodic lattice. For a square array with
frustration $f$ and with the random displacements in $x$ and $y$
directions distributed uniformly in the range $[ - \Delta ,\Delta
]$ with $\Delta $ in units of the lattice constant, the
corresponding disorder strength is given by the relation $r
\approx 2.2f\Delta $ \cite{R10,R11}. Since the $XY$ Hamiltonian
is invariant under the transformation of $f$ to $f + n$ for an
integer $n$, the disorder strength of a JJA for a specific $f$ can
be controlled by changing either $n$ or $\Delta $. In this paper,
we present an experimental investigation of a positionally
disordered JJA with $f = n$, $\frac{1}{2} + n$, or $\frac{2}{5} +
n$ for various $n$'s. Transport measurements revealed that the
three different systems superconduct at sufficiently low
temperatures for any disorder strength $r$ and, despite the
essential differences among them at weak disorder, belong to the
same universality class in the limit of maximal disorder,
proposing the possible occurrence of a finite-temperature glass
transition. For $f = 0$ and $\frac{1}{2}$, the nature of the
superconducting transition was found to change from a KT type to
a non-KT type with $r$ increased.

Experiments were performed on a square array of 200$ \times $800
Nb/Cu/Nb Josephson junctions with a controlled amount positional
disorder, a section of which is shown in Fig.\ \ref{fig1}(a). Nb
islands were disposed on a 0.3-$\mu $m-thick Cu film periodically
with a lattice constant of 13.7 $\mu $m, a junction width of 4
$\mu $m, and a junction separation of 1.4 $\mu $m. The random
displacements of the centers of the crosses in $x$ and $y$
directions were uniformly distributed in the range $[ - \Delta
,\Delta ]$ with $\Delta  = 0.20$ in units of the lattice constant.
For an ideal JJA with zero junction width, this amount of
displacement corresponds to the disorder strength $r \approx
0.44f$. However, the actual disorder strength for our sample with
a finite junction width must be weaker than designed, due to the
asymmetric diamagnetic responses of the distorted superconducting
islands which inhomogeneously changes the effective area of each
cell to reduce the amount of positional disorder \cite{R12}. If
the relocation of the effective center of the Nb island is
limited within the middle square of the cross, the smallest
possible value of the effective displacement $\Delta _{eff} $ for
$\Delta = 0.20$ is $\sim 0.05$. The actual amount of $\Delta
_{eff} $ of the sample will be determined in the stage of
comparing the experimental results with the numerically proposed
phase diagram. In the absence of a magnetic field, $i_c $ and $J$
($ = \hbar i_c /2e$) can be obtained by extrapolating the
low-temperature $i_c $ vs $T$ data, obtained from the $I$ vs
$dV/dI$ curves, by using de Gennes formula \cite{R13} for
proximity-coupled junctions in the dirty limit. In the presence
of a magnetic field, $i_c $ depends on the magnetic flux $\phi $
through the junction area as $i_c (B) = i_c (0)\sin (\pi \phi
/\phi _0 )/(\pi \phi /\phi _0 )$. Since it is not possible to
directly measure $i_c (B)$ of a positionally disordered array and
$i_c (B)/i_c (0)$ ($ = \sin (\pi \phi /\phi _0 )/(\pi \phi /\phi
_0 )$) depends mainly on the geometry of the junction, $i_c (B)$
of the sample was determined by combining $i_c (0)$ of the sample
and $i_c (B)/i_c (0)$ ($ = \sin (\pi \phi /\phi _0 )/(\pi \phi
/\phi _0 )$) of another array with the same junction dimensions
and properties but no positional disorder. The resistance ($R$)
and the current-voltage ($IV$) characteristics were measured by
employing the phase-sensitive voltage-signal-detection method
using a lock-in voltmeter and a square-wave current at 23 Hz. The
magnetic field or the frustration $f$ was adjusted from the $f$
vs $R$ curve of the sample exhibiting distinct resistance minima
at integral $f$'s. More details of the experiments are described
in Ref.\ \cite{R14}.

Figure\ \ref{fig1}(b) shows the resistive transition of the
sample for three different sets of  frustrations $f = n$,
$\frac{1}{2} + n$, and $\frac{2}{5} + n$. Although the transition
broadens appreciably with increasing $n$, it is evident that for
all the frustrations, the transition to a zero-resistance state
takes a place at a finite temperature. Figure\ \ref{fig2}
illustrates the $f$ dependence of the superconducting transition
temperature $T_c $ determined at $R = 5 \times 10^{-6}$ $\Omega$.
Note that $T_c $ is plotted in units of $J/k_B $ in order to
scale out the temperature and magnetic-field dependences of the
critical current $i_c (T,B)$ and to make a direct comparison of
the $T_c $ with the numerical results possible. The $T_c $'s
determined from the resistance measurements agree within error
bars with those independently determined from the analysis of the
$IV$ characteristics (to be commented on in detail below). We
first consider the $f = 0$ case.  As $n$ (or the disorder
strength) increases, $T_c $ decreases at first from 0.84$J/k_B $
and then stays at $\sim 0.35J/k_B $ for $f > f_c \sim 3$. The
phase line never ends at $T = 0$, contrary to many earlier
investigations \cite{R2,R3}. A similar phase boundary with $f_c
\sim 5$ was observed for the sample with $\Delta  = 0.15$. The
nature of the superconducting transition could be identified from
the development of the $IV$ characteristics with temperature.
Measurements of the $IV$ characteristics were carried out for six
different frustrations $f = $ 0, 2, 3, 4, 6, and 8. Shown in
Fig.\ \ref{fig3} are those for $f = $ 2 and 6. The $IV$ curves
were obtained by averaging 15-240 measurements for each current.
For $f = 2$, the $IV$ curves below the transition display a
power-law behavior (a straight line in a log$I$ vs log$V$ plot or
a flatted curve in a log$I$ vs $d(\log V)/d(\log I)$ plot) at low
currents, indicative of the KT-type order with a quasi-long-range
phase coherence and vortex-antivortex pairs as dominant
excitations. The fast drop of the slope $d(\log V)/d(\log I)$ of
the $IV$ curves with temperature above the transition is also
characteristic of the KT transition. The $IV$ curves for $f = 6$,
on the other hand, have an exponential form (a convex upward
curve in a log-log plot) below the transition, indicating a
non-KT type order with a long-range phase coherence. As shown at
the bottom of Fig.\ \ref{fig3}, the $IV$ curves for $f = 6$ can
be fitted to the Fisher-Fisher-Huse scaling form \cite{R1} for a
non-KT type superconducting transition in 2D, $TV/I|T - T_c
|^{z\nu }  = \varepsilon _ \pm (I/T|T - T_c |^\nu )$, where $z$
is the dynamic exponent and $\nu $ is the correlation length
exponent \cite{R15}.  The critical exponents from the scaling
analysis are $\nu  = 2.0 \pm 0.3$ and $z = 1.8 \pm 0.3$ for $f >
3$, insensitive to $f$. Crossing over from the KT-like behavior
to the non-KT-like one arose at $f \sim f_c $ ($\sim 3$), the
frustration above which $T_c $ became approximately independent
of $f$. The evolution of the $IV$ characteristics with $f$
discloses that the nature of the superconducting transition for
$f = 0$ changes from a KT type at weak disorder to a non-KT type
at strong disorder. These experimental findings except the values
of $T_c $ and $\nu $ at strong disorder are in a good agreement
with the numerical results reported in Ref.\ \cite{R5}. Comparing
the phase boundary for $f = 0$ in Fig.\ \ref{fig2} with the
numerically proposed phase diagram in Ref.\ \cite{R5}, we find
that $f_c \sim 3$ for our sample corresponds to the critical
disorder strength $r_c  \approx 0.37$ and thus $f \sim 8$ to $r =
1$. This implies a finite-temperature superconducting transition
even in the gauge glass limit of $r = 1$. The correspondence of
$f_c \sim 3$ with $r_c  \approx 0.37$ also gives $\sim 0.06$ for
$\Delta _{eff} $ of the sample.

The effect of positional disorder on the $f = \frac{1}{2}$ system
which has the KT-like phase coherence and the Ising-like
structural order in the absence of disorder were found quite
similar to the observed for $f = 0$. As shown in Fig.\
\ref{fig2}, with increasing $n$, $T_c $ of the $f = \frac{1}{2}$
system drops fast from 0.45$J/k_B $ (the $T_c $ in the absence of
disorder) to $\sim 0.25J/k_B $ and then increases slowly until it
reaches $\sim 0.3J/k_B $. The $IV$ characteristics for $f =
\frac{1}{2}$, $1\frac{1}{2}$, $3\frac{1}{2}$, $5\frac{1}{2}$, and
$7\frac{1}{2}$, some of which are shown in figures\ \ref{fig4}
and \ \ref{fig5}(b), indicate that similarly to the $f = 0$ case,
the nature of the superconducting transition changes from a KT
type at weak disorder to a non-KT type at strong disorder. The
critical exponents for the non-KT type transitions are $\nu  =
2.0 \pm 0.3$ and $z = 1.8 \pm 0.3$. The major difference from the
$f = 0$ case is that the crossing over appears at lower $f_c
\lesssim 1.5$.

Dissimilarly from the $f = 0$ and $\frac{1}{2}$ systems, the $f =
\frac{2}{5}$ system without disorder has a superconducting
vortex-solid phase with a long-range phase coherence at low
temperatures and experiences a melting transition driven by
domain wall excitations \cite{R16}.  The $IV$ characteristics for
$f = \frac{2}{5}$, $1\frac{2}{5}$, $3\frac{2}{5}$,
$5\frac{2}{5}$, and $7\frac{2}{5}$, some of which are shown in
figures\ \ref{fig4} and \ \ref{fig5}(b), reveal that for $f =
\frac{2}{5}$, the introduction of positional disorder does not
alter the essential features of the low-temperature ordered state
and phase transition. The phase diagram of Fig.\ \ref{fig2}
displays $T_c $ for $f = \frac{2}{5}$ as a monotonously
increasing function of the disorder strength. The scaling
analyses of the $IV$ data present 1.8-2.0 for $\nu $ and 1-1.8
for $z$, dependent on the disorder strength.

The experimental results for the three different sets of
frustrations demonstrate that as the disorder strength (or the
amount of the random phase shifts) increases, a KT type order of
a 2D $XY$ system weakens and is eventually replaced by a non-KT
type order with a long-range phase coherence, which never weakens
even at the maximum disorder strength. A remaining important
question to be answered through this work is whether the
finite-temperature superconducting phase in the maximal disorder
limit is a glass in nature. Figure\ \ref{fig5}(a) exhibits the
variation of the sample resistance with $f$ at $T = 4.20$ K. The
resistance oscillation fades away as $f$ approaches the maximal
disorder limit ($f \sim 8$). Three sets of $IV$ curves for $f =
8$, $7\frac{1}{2}$, and $7\frac{2}{5}$ plotted in a single figure
[Fig.\ \ref{fig5}(b)] also shows that the $IV$ characteristics
near maximal disorder are independent of frustration, excluding
the slightly $f$-dependent $T_c $. Consequently, the three sets of
$IV$ data collapse onto the same curves with $I$ and $V$ scaled
by the Fisher-Fisher-Huse scaling form with the same $\nu $ ($=
2.0$) and $z$ ($= 1.85$).  This suggests that the 2D $XY$ systems
with the maximal disorder strength belong to the same
universality class, regardless of the amount of frustration
imposed on the system, or that the natures of the superconducting
state and phase transition of a 2D $XY$ system become independent
of $f$ in the gauge glass limit, despite the essential
differences among the systems with different $f$'s at weak
disorder. Although the data do not provide a full account for the
nature of the low-temperature superconducting state, these
features of the data seem to propose the possible occurrence of a
finite-temperature glass transition in the gauge glass limit.

\newpage

\begin{figure}[h!]
\caption{(a) Photograph of the positionally disordered array with
$\Delta  = 0.20$. (b) Resistive transition of the disordered
array for $f = n$, $\frac{1}{2} + n$, and $\frac{2}{5} + n$.}
\label{fig1}
\end{figure}

\begin{figure}[h!]
\caption{Superconducting transition temperature $T_c$ in units of
$J/k_B $ for three different sets of frustrations $f = n$ (filled
circles), $\frac{1}{2} +n$ (open squares), and $\frac{2}{5} + n$
(filled triangles), determined from the resistance measurements.}
\label{fig2}
\end{figure}

\begin{figure}[h!]
\caption{$I$ vs $V$ curves in a log-log scale for $f = 2$ at $T =
$ 4.200 to 5.200 K and for $f = 6$ at $T = $ 3.800 to 4.900 K; the
slope $d(\log V)/d(\log I)$ of the $I$ vs $V$ curves as a
function of $I$; and scaling plots of the $I$ vs $V$ curves at
12-18 different temperatures. The dashed lines in the $I$ vs $V$
plots are drawn to show where the phase transition occurs. The
insets of the scaling plots show the values of $T_c $, $\nu $,
and $z$ used to scale the data.} \label{fig3}
\end{figure}

\begin{figure}[h!]
\caption{Development of $I$ vs $V$ curves with temperature for $f
= \frac{1}{2}$, $3\frac{1}{2}$, $\frac{2}{5}$, and
$3\frac{2}{5}$. The dashed lines are drawn to show where the
phase transition occurs.} \label{fig4}
\end{figure}

\begin{figure}[h!]
\caption{(a) Frustration dependence of the resistance of the
sample with an excitation current of 30 $\mu $A at $T = 4.20$ K.
(b) $I$ vs $V$ curves for three different frustrations $f = 8$,
$7\frac{1}{2}$, and $7\frac{2}{5}$ in a single plot.} \label{fig5}
\end{figure}


\begin{references}
\bibitem {R15}
Although there exists some doubt about the validity of the
scaling analysis of $IV$ data for high-temperature
superconductors, it has been also shown that this does not have
to be the case for such a large array as ours, when the $IV$ data
are taken by employing the phase-sensitive
voltage-signal-detection method. Detailed discussions on this
matter are given in Ref.\ \cite{R14} and references therein.
\end{references}
\end{document}